\documentclass[prl,twocolumn]{revtex4}
\usepackage{amssymb,amsmath,amsthm,graphicx}
\usepackage{latexsym}
\usepackage{bm}

\def\be{\begin{equation}}
\def\ee{\end{equation}}
\def\beq{\begin{eqnarray}}
\def\eeq{\end{eqnarray}}

\pdfoutput=1

\begin{document}

\def\lsim{\mathrel{\rlap{\lower4pt\hbox{\hskip1pt$\sim$}}
    \raise1pt\hbox{$<$}}}
\def\gsim{\mathrel{\rlap{\lower4pt\hbox{\hskip1pt$\sim$}}
    \raise1pt\hbox{$>$}}}
%\def\sqr#1#2{{\vcenter{\vbox{\hrule height.#2pt
%         \hbox{\vrule width.#2pt height#1pt \kern#1pt
%         \vrule width.#2pt}
%         \hrule height.#2pt}}}}
%\def\square{\mathchoice\sqr66\sqr66\sqr{2.1}3\sqr{1.5}3}
%%%%%%%%%%%%%%%%%%%%%%%%%%%%%%
\def\be{\begin{equation}}
\def\ee{\end{equation}}
\def\bea{\begin{eqnarray}}
\def\eea{\end{eqnarray}}
%%%%%%%%%%%%%%%%%%%%%%%%%%%%%%
\newcommand{\dd}{\mathrm{d}}
\newcommand{\LL}{\mathcal{L}}
\newcommand{\DD}{\mathcal{D}}

\title{Linear mode stability of Kerr-Newman and its quasinormal modes}
%\title{Lumpy black holes: connecting black holes to black rings}

\author{\'Oscar J.~C.~Dias$^{a}$, Mahdi Godazgar$^{b}$,  Jorge E.~Santos$^{b}$\,}
\email{ojcd1r13@soton.ac.uk, mmg31@damtp.cam.ac.uk, jss55@cam.ac.uk}
\affiliation{$\,$\\$^{a}$STAG research centre and Mathematical Sciences, University of Southampton, UK\\$\,$\\
$^{b}$DAMTP, Centre for Mathematical Sciences,
    University of Cambridge, Wilberforce Road, Cambridge CB3 0WA, UK}

\begin{abstract}

We provide strong evidence that, up to $99.999\%$ of extremality, Kerr-Newman black holes (KN BHs) are linear mode stable within Einstein-Maxwell theory. We derive and solve, numerically, a coupled system of two PDEs for two gauge invariant fields that describe the most general linear perturbations of a KN BH (except for trivial modes that shift the parameters of the solution). We determine the quasinormal mode (QNM) spectrum of the KN BH as a function of its three parameters and find no unstable modes. In addition, we find that the QNMs that are connected continuously to the gravitational $\ell=m=2$ Schwarzschild QNM dominate the spectrum for all values of the parameter space ($m$ is the azimuthal number of the wave function and $\ell$ measures the number of nodes along the polar direction). Furthermore, all QNMs with $\ell=m$ approach Re$\,\omega = m \Omega_H^{ext}$ and Im$\,\omega=0$ at extremality; this is a universal property for any field of arbitrary spin $|s|\leq 2$ propagating on a KN BH background ($\omega$ is the wave frequency and $\Omega_H^{ext}$ the BH angular velocity at extremality). We compare our results with available perturbative results in the small charge or small rotation regimes and find good agreement.  We also present a simple proof that the Regge-Wheeler (odd) and Zerilli (even) sectors of Schwarzschild perturbations must be isospectral. 
\end{abstract}

\maketitle

%%%%%%%%%%%%%%%%%%%%%%%%%%%%%%%%%%%%%%%%%%%%%%%%%%%%%%%%%%%%%%%%%%%%%%%%%%%
%%%%%%%%%%%%%%%%%%%%%%%%%%%%%%%%%%%%%%%%%%%%%%%%%%%%%%%%%%%%%%%%%%%%%%%%%%%

%%%%%%%%%%%%%%%%%%%%%%%%
\emph{\bf  Introduction.}
%%%%%%%%%%%%%%%%%%%%%%%%
The uniqueness theorems \cite{Robinson:2004zz,Chrusciel:2012jk} state that the Kerr-Newman black hole (KN BH) \cite{Newman:1965my,Adamo:2014baa} is the unique, most general family of stationary asymptotically flat BHs, of Einstein-Maxwell theory. It is characterised by 3 parameters: mass $M$, angular momentum $J\equiv M a$ and charge $Q$. The Kerr, Reissner-Nordstr\"om (RN) and Schwarzschild (Schw) BHs constitute limiting cases: ${Q=0}$, $a=0$ and $Q=a=0$, respectively.

Given their uniqueness, the most relevant question to consider is the linear mode stability of these BHs. It is known that the Kerr, RN, and Schw BHs are linear mode stable. Indeed, the perturbation study of the linearised Einstein(-Maxwell) equation gives the quasinormal mode (QNM) spectrum of frequencies (that describes the damped oscillations of the BH back to equilibrium) resulting in no unstable modes \cite{Regge:1957td,Zerilli:1974ai,Moncrief:1974am,Chandrasekhar:1975zza,Moncrief:1974gw,Moncrief:1974ng,Newman:1961qr,Geroch:1973am,Teukolsky:1972my,Detweiler:1980gk,Chandra:1983,Leaver:1985ax,Whiting:1988vc,Onozawa:1996ux,Yang:2012pj,Berti:2003jh} (see review \cite{Berti:2009kk}). Remarkably, for these BHs, the QNM spectrum turns out be encoded in a single separable equation---known as the Regge-Wheeler--Zerilli equation \cite{Regge:1957td,Zerilli:1974ai,Moncrief:1974am} (for RN and Schw) and the Teukolsky equation \cite{Teukolsky:1972my} (for Kerr, RN and Schw)---that effectively yields a pair of ODEs.

Unfortunately, it does not seem possible to cast a general perturbation of a KN BH as a single PDE. Therefore, obtaining the QNM spectrum of KN BHs requires solving coupled PDEs. Na\"ively, one expects to find a system of nine coupled PDEs. However, working in the so-called phantom gauge, $\Phi_{0}^{(1)}=\Phi_{1}^{(1)}=0$, Chandrasekhar, reduced the problem to the study of two coupled PDEs \cite{Chandra:1983} (see also \cite{Mark:2014aja}). Despite this significant progress, finding the QNM spectrum and addressing the problem of the linear mode stability of the KN BH has remained a major open problem of Einstein-Maxwell theory since the 80's, when Chandrasekhar's seminal work \cite{Chandra:1983} was published.

Recently there have been some notable efforts to address this problem. Refs. \cite{Pani:2013ija,Pani:2013wsa} and \cite{Mark:2014aja} have found the QNM spectrum in a perturbative small rotation and charge, respectively, expansion around the RN and Kerr BHs. These works find no sign of linear instability; however such an instability is more likely to be found in extreme regimes where both $Q$ and $a$ are large. Another remarkable effort to infer the (non-)linear stability of KN BHs has been made in \cite{Zilhao:2014wqa}, where the full time evolution of some KN BH with a given initial perturbation is considered, finding no sign of a nonlinear instability. However, since non-linear simulations are computationally costly, the search in moduli space is modest.

In this letter, we derive two coupled PDEs that reduce to the Chandrasekhar coupled PDE system upon gauge fixing and compute the QNM spectrum of the KN BH to a high degree of accuracy. Up to $a/a_{ext}=0.99999$ we find no sign of a linear mode instability for any of the gravito-electromagnetic modes that are described by $\ell=1,2,3,4$ and $|m|\leq \ell$. We use two distinct numerical methods that have been developed to solve efficiently similar systems of (several coupled) ODEs and PDEs that appear in QNM, superradiant and ultraspinning instability studies \cite{Dias:2009iu,Dias:2010eu,Dias:2010maa,Dias:2010gk,Dias:2011jg,Dias:2010ma,Dias:2011tj,Dias:2013sdc,Cardoso:2013pza,Dias:2014eua}. One of these methods formulates the problem as a quadratic eigenvalue problem in the frequency and employs a pseudospectral grid collocation. The other method searches directly for specific QNMs using a Newton-Raphson root-finding algorithm. We refer the reader to \cite{Dias:2009iu,Dias:2010eu,Dias:2010maa,Dias:2010gk,Dias:2011jg,Dias:2010ma,Dias:2011tj,Dias:2013sdc,Cardoso:2013pza,Dias:2014eua} for details. The pseudospectral exponential convergence of our method, and the use of quadruple precision, guarantees that the results are accurate up to, at least, one tenth of a decimal place.

\emph{Notation:} We use the standard nomenclature of the Newman-Penrose formalism to denote components of the curvatures, electromagnetic field strength and connections \cite{Stephani:2003tm}. $X^{(0)}$ denotes background quantities, while $X^{(1)}$ denotes a perturbed quantity at the linear order.

%%%%%%%%%%%%%%%%%%%%%%%%%
\emph{\bf  Formulation of the problem.}
%%%%%%%%%%%%%%%%
We write the KN BH solution in standard Boyer-Lindquist coordinates $\{t,r,\theta,\phi\}$ (time, radial, polar, azimuthal coordinates) \cite{Adamo:2014baa}. Its event horizon, with angular velocity $\Omega_H$ and temperature $T_H$, is generated by the Killing vector $K=\partial_t +\Omega_H \partial \phi$. The location of the horizon $r_+$ is given by the largest root of the function $\Delta$. These quantities are given in terms of the parameters $\{M,a,Q\}$ as follows:  
\begin{eqnarray}\label{KNsol}
&& \Delta = r^2 -2Mr+a^2+Q^2,\ \  r_+=M+\sqrt{M^2-a^2-Q^2}, \nonumber \\
&& \Omega_H= \frac{a}{r_+^2+a^2} \,, \qquad 
T_H = \frac{1}{4 \pi  r_+}\frac{r_+^2-a^2-Q^2}{r_+^2+a^2 }.
\end{eqnarray}
The KN BH has a regular extremal configuration when its temperature vanishes and its angular velocity reaches a maximum. For fixed $M$ and $Q$ this occurs for $a=a_{ext}=\sqrt{M^2-Q^2}$.  Thus, at extremality we have 
\begin{equation}\label{KNsolExt}
T_H^{ext} =0 \quad \Leftrightarrow \quad
\Omega_H^{ext} = \frac{\sqrt{M^2-Q^2}}{2 M^2-Q^2}=\frac{a}{M^2+a^2}\,.
\end{equation}

We consider the most general perturbation of a KN BH. Using the fact that $\partial_t$ and $\partial_\phi$ are Killing vector fields of the KN background, we Fourier decompose its perturbations as $e^{-i \omega t} e^{i m \phi}$. This introduces the frequency $\omega$ and azimuthal quantum number $m$ of the perturbation. By formulating the perturbation problem in the Newman-Penrose (NP) formalism, we obtain a set of two coupled partial differential equations that describe the most general linear perturbation of a KN BH (see Appendix for details of the derivation)
\begin{eqnarray} \label{coupledeqns}
&& \left(\mathcal{O}_{-2} + \Phi_{11}^{(0)} \mathcal{P}_{-2}\right) \varphi_{-2} + \Phi_{11}^{(0)} \mathcal{Q}_{-2} \varphi_{-1} =0 \,, \\
&& \left(\mathcal{O}_{-1} + \Phi_{11}^{(0)} \mathcal{P}_{-1}\right) \varphi_{-1} + \Phi_{11}^{(0)}  \mathcal{Q}_{-1} \varphi_{-2}=0  \,, \nonumber
\end{eqnarray}
where differential operators $\{\mathcal{O},\mathcal{P},\mathcal{Q}\}$ are given in \eqref{def:opsOPQ} of Appendix, $\varphi_{-2}=\Psi_4^{(1)}$ and ${\varphi_{-1} = 2 \Phi_1^{(0)} \Psi_3^{(1)} - 3 \Psi_2^{(0)} \Phi_{2}^{(1)}}.$

Substituting the background values of the NP quantities, the above equations reduce to
\begin{eqnarray}\label{ChandraEqs}
&& \left(\mathcal{F}_{-2}+ Q^2 \mathcal{G}_{-2}\right) \psi_{-2}  + Q^2 \mathcal{H}_{-2}  \psi_{-1} =0 \,, \\
&& \left(\mathcal{F}_{-1} +Q^2 \mathcal{G}_{-1}\right)\psi_{-1} + Q^2  \mathcal{H}_{-1} \psi_{-2}=0  \,, \nonumber
\end{eqnarray}
where second order differential operators $\{\mathcal{F},\mathcal{G},\mathcal{H}\}$ are given in \eqref{def:opsFGH} of Appendix and
\begin{eqnarray}\label{gauging}
&&\psi_{-2}= \left(\bar{r}^*\right)^4 \Psi_4^{(1)},  \nonumber\\
&& \psi_{-1}=\frac{\left(\bar{r}^*\right)^3}{2\sqrt{2}\Phi_1^{(0)}} \left(2\Phi_1^{(0)}\Psi_3^{(1)} -3 \Psi_2^{(0)}\Phi_2^{(1)}\right) \,
\end{eqnarray}
with $\bar{r} = r+ia\cos \theta$.  We emphasise that $\psi_{-2}$ and $\psi_{-1}$ (as well as $\varphi_{-2}$ and $\varphi_{-1}$) are {\it gauge invariant} perturbed quantities, i.e.\ they are invariant under both linear diffeomorphisms and tetrad rotations.  Furthermore, these are the NP scalars that are relevant for the study of perturbations that are outgoing at future null infinity and regular at the future horizon \footnote{There is a set of two coupled PDEs---related to (4) by a Geroch-Held-Penrose \cite{Geroch:1973am} transformation---for the quantities $\psi_{2}$ and $\psi_{1}$ that are the positive spin counterparts of (4); however these would be relevant if we were interested in perturbations that were outgoing at past null infinity.}.  Fixing a gauge in which $\Phi_{0}^{(1)}=\Phi_{1}^{(1)}=0$, we obtain the Chandrasekhar coupled PDE system \cite{Chandra:1983} (see also the derivation in \cite{Mark:2014aja}).  Finally, note that in the limit $Q\to 0$ and/or $a\to0$ these equations decouple yielding the Teukolsky equation.

In order to solve these equations, we need to impose appropriate boundary conditions (BCs). The $t -\phi$ symmetry of the KN BH guarantees that we can  consider only modes with $m\geq 0$, say, as long as we consider both positive and negative Re$(\omega)$; when $a=0$ this enhances to a $t \to -t $ symmetry and the QNM frequencies form pairs of $\{\omega,-\omega^* \}$. 

At spatial infinity, a Frobenius analysis of \eqref{ChandraEqs} and the requirement that we have only outgoing waves fixes the decay to be ($s=-2,-1$)
\begin{equation}\label{BC:inf}
\psi_{s}{\bigl |}_\infty \!\simeq\!e^{i \omega r} r^{-(2s+1)+i \omega \,\frac{r_+^2+a^2+Q^2}{r_+}} \!\! \left( \!\! \alpha_{s}(\theta)+\frac{\beta_{s}(\theta)}{r}+\cdots\!\!\right), \nonumber
\end{equation}
where $\alpha_{s}(\theta)$ and $\beta_{s}(\theta)$ are related by two Robin BCs which are fixed by \eqref{ChandraEqs}. 

At the horizon, a Frobenius analysis and requiring only regular modes in the ingoing Eddington-Finkelstein coordinates yields the near-horizon expansion 
\begin{equation}\label{BC:H}
\psi_{s}{\bigl |}_H \!\simeq\! \left(r-r_+\right)^{-s-\frac{i (\omega -m\Omega_H)}{4 \pi T_H}}[ a_{s}(\theta)+ b_{s}(\theta)(r-r_+) +\cdots ], \nonumber
\end{equation}
where $b_{s}(\theta)$ is related to $a_s(\theta)$ and its derivatives.

At the North (South) pole $x\equiv \cos\theta =1\,(-1)$, regularity dictates that the fields behave as 
($\varepsilon = 1$ for $m\geq 2$, while $\varepsilon = -1$ for $m=0,1$ modes)
\begin{equation}\label{BC:N}
\psi_{s}{\bigl |}_{N,(S)} \hspace{-1pt} \simeq  (1\mp x)^{\varepsilon^{\frac{1\pm 1}{2}} \frac{s+m}{2}} \hspace{-1pt} \left[ A^{\pm}_{s}(r)+B^{\pm}_{s}(r)(1\mp x)+\cdots \right],  \nonumber
\end{equation}
where $A^{\pm}_{s}(r)$ and $B^{\pm}_{s}(r)$ are related by two Robin BCs that are fixed by the equations of motion.

We consider only modes with the lowest radial overtone ($n=0$) because these are the ones that have smaller $|\mathrm{Im}\,\omega|$ and thus they are the ones that dominate a time evolution and are more likely to become unstable near extremality. Note also that we can scale out one of the 3 parameters of the solution. Thus, we work with the adimensional parameters $\{a/M,Q/M\}$ (or $\{a/r_+,Q/r_+\}$) and $\omega M$.

\begin{figure}[th]
\centering
\includegraphics[width=.42\textwidth]{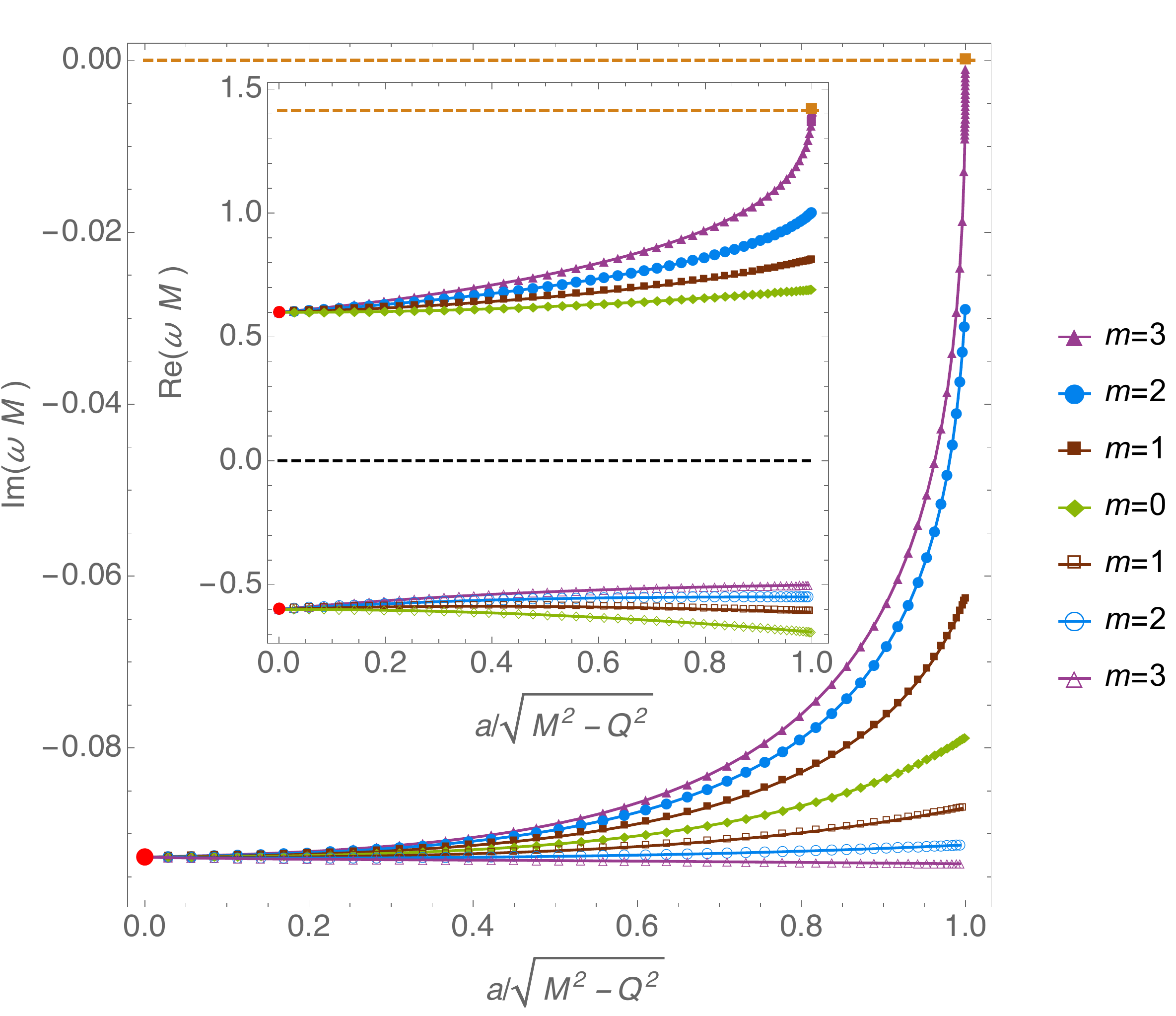}
\caption{All lowest radial overtone QNMs of $Q=a$ KN BHs that start at the $\ell=3$ Schw gravitational QNM (red disc).}\label{Fig:qaz2l3}
\end{figure}  
%%%%%%%%%%%%%%%%%%%%%%%%
\emph{\bf Results and Discussion.}
%%%%%%%%%%%%%%%%%%%%%%%%
Our primary aim is to find whether KN BHs can be linear mode unstable. For that, we study the QNM spectrum and check if there are modes with Im$\,\omega>0$.  Note that for $Q,a\to 0$ we ought to recover the Schw QNMs. In this limit, there are two families of QNMs, namely the Regge-Wheeler (odd or axial) modes and the Zerilli (even or polar) modes. These families are isospectral, i.e. they have exactly the same spectrum (\cite{Chandra:1983}; more later). Thus we only need  to distinguish the gravitational modes (described in Table V of page 262 \cite{Chandra:1983}---hereafter Table of \cite{Chandra:1983}---by the eigenfunction $Z_2$) from the electromagnetic modes (described in Table of \cite{Chandra:1983} by the eigenfunction $Z_1$). Each of these is specified by the harmonic number $\ell=1,2,3,\cdots$ ($Z_2$ modes with $\ell=1$ are pure gauge modes). When the BH has charge and rotation, we have to scan a two parameter space in $\{Q/M,a/M\}$. The above two families become coupled gravito-electromagnetic QNMs and the Schwarzschild eigenvalue $\ell$ does not appear explicitly in the KN PDEs \eqref{ChandraEqs}. However, we can still count the number of nodes along the polar direction of the eigenfunctions of \eqref{ChandraEqs} and this gives $\ell$. 

We perform a {\it complete} scan in $\{Q/M,a/M\}$ for {\it all} modes with $\ell=1,2,3$, $|m|\leq3$ (both in the $Z_1$ and $Z_2$ sectors). Modes with $\ell = 4$ are also studied, but there we focus on modes that approached Im$\,\omega =0$ at extremality. As one of our main results, we do {\it not} find any unstable mode with Im$\,\omega>0$, even when we probe regions in parameter space for which $a/a_{ext} = 0.99999$. We see this as good numerical evidence that the KN BH is linearly mode stable.

\begin{figure}[th]
\centering
\includegraphics[width=.45\textwidth]{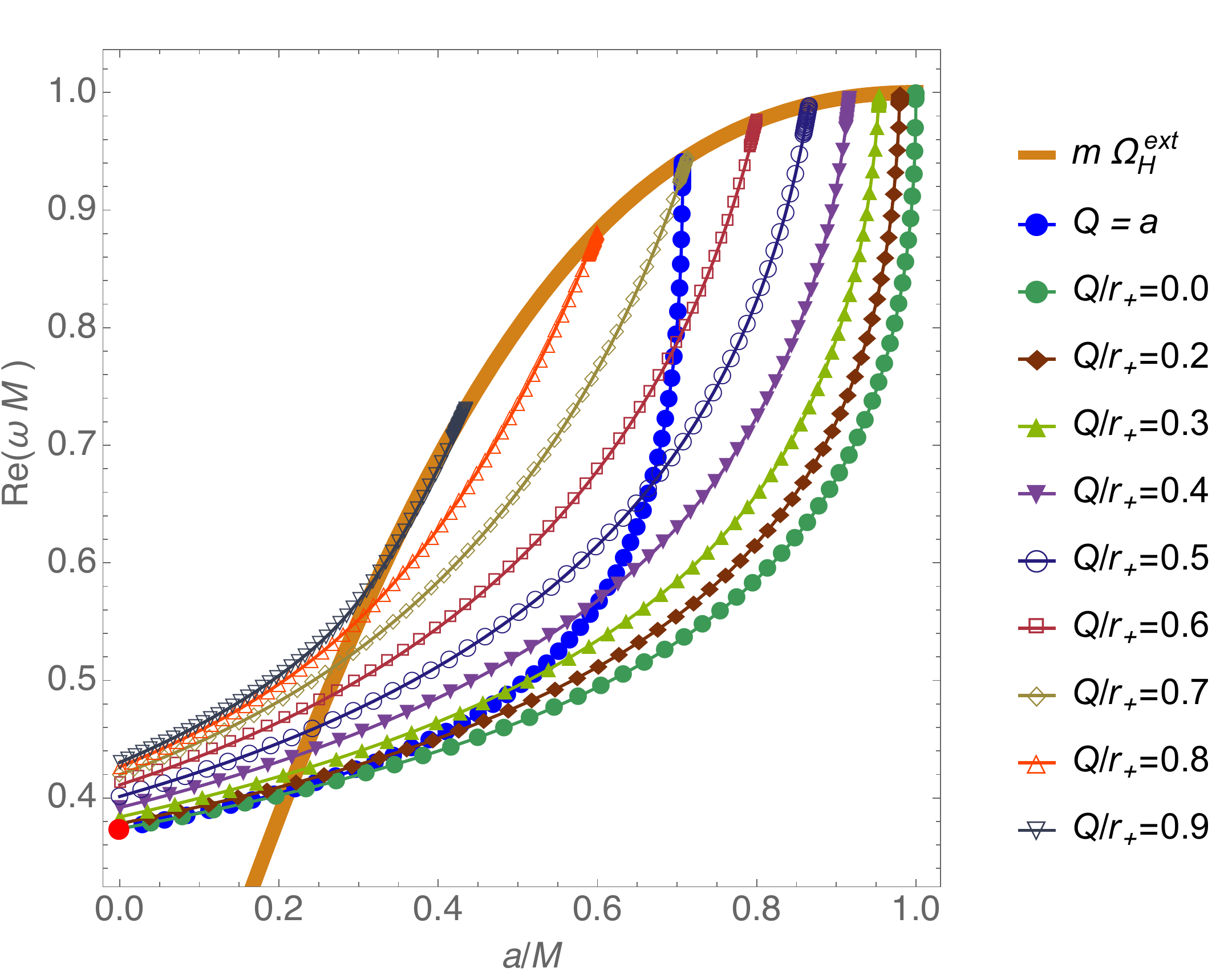}
\caption{Real part of the QNM frequencies with $\ell=m=2$  of KN BHs with $Q=a$ and fixed $Q/r_+=0.0,0.2,\cdots,0.9$ that start at the Schw gravitational QNM with $\ell=2$ (red disc).}\label{Fig:l2m2re}
\end{figure}  

\begin{figure}[th]
\centering
\includegraphics[width=.45\textwidth]{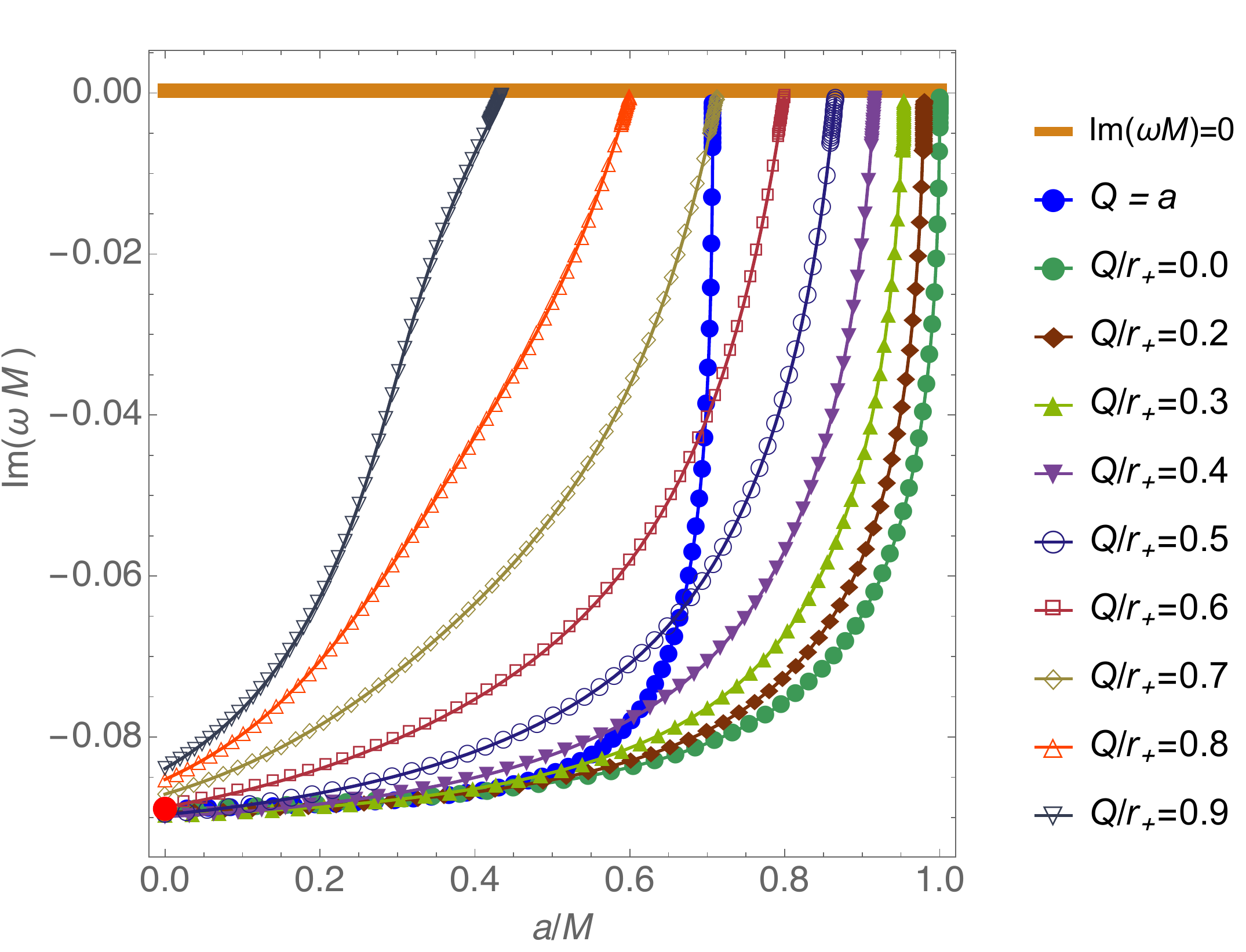}
\caption{Similar to Fig. \ref{Fig:l2m2re} but now for Im$(\omega M)$.}\label{Fig:l2m2im}
\end{figure}  

To illustrate our search, in Fig. \ref{Fig:qaz2l3} we take KN BHs with $Q=a$ and we display all the QNMs that are continuously connected to the gravitational $Z_2$ Schw QNM with $\ell=3$, namely $\omega M=0.59944329- 0.09270305\, i $ (see Table of \cite{Chandra:1983}). The different QNMs are distinguished  by their azimuthal number $m=0,1,2,3$ and by whether they have positive or negative Re$\,\omega$ (modes with $m=0$ have a pair of QNMs $\{\omega,-\omega^* \}$; see discussion above). All these modes become degenerate in the  Schwarzschild limit (red disk in Fig. \ref{Fig:qaz2l3}). We plot the imaginary (main plot) and real (inset plot) parts of the frequency $\omega M$ as a function of $a/a_{ext}=a/\sqrt{M^2-Q^2}$. We see that the most likely mode to be unstable is the $\ell = 3$ mode with Re$\,\omega>0$ (magenta triangles). However, we follow this mode up to $a/a_{ext}=0.99999$ and find that although the Im$\,\omega$ quickly approaches zero as $a\to a_{ext}$, it never crosses Im$\,\omega =0$.

It is also relevant to ask what are the dominant QNMs, i.e. the modes with the smallest $|\mathrm{Im}\,\omega|$. We find that the QNM family that, in the $Q,a\to 0$ limit, approaches the $Z_2$ Schw QNM with $\ell=m=2$ with $\omega M=0.37367168 -0.08896232\, i $  (Table of \cite{Chandra:1983}) is the one that {\it always} (i.e. for a given $Q$ and $a$) has the smallest $|\mathrm{Im}\,\omega|$. Therefore, these QNMs must be the dominant modes in a time evolution of the KN BH. Since this mode is the most relevant in a time evolution process, hereafter we will use it to illustrate our discussions (other modes will be presented elsewhere). 

The plots of Figs. \ref{Fig:l2m2re} (real part of $\omega M$) and  \ref{Fig:l2m2im}  (imaginary part of $\omega M$) give details of the $Z_2, \ell=m=2$ mode. We represent the QNMs of the  $Q=a$ KN BH by blue disks, but we also present the QNMs for KN BHs with fixed charge $Q/r_+$ (see plot legends)  as the rotation grows from zero to $a_{ext}/M$. Fig. \ref{Fig:l2m2re} shows that as extremality is approached we always have Re$\,\omega\to 2\Omega_H^{ext}$. On the other hand, Fig. \ref{Fig:l2m2im} shows that Im$\,\omega \to 0^{-}$ as extremality is approached . Again, we emphasise that the last point of each of these curves is, at least, at $a=0.99999\,a_{ext}$.  
Figs. \ref{Fig:qaz2l3}, \ref{Fig:l2m2re} and \ref{Fig:l2m2im} illustrate a general property of the KN QNMs with $\ell=m$ (both in the $Z_1$ and $Z_2$ sectors): as extremality is approached one has Re$\,\omega\to m \,\Omega_H^{ext}$ and Im$\,\omega\to 0^{-}$. Collecting previous results \cite{Detweiler:1980gk,Glampedakis:2001js,Berti:2003jh,Hod:2008zz},  we can can now state that this is a universal property for any perturbation spin $s$. As we discussed above, in a 3-dimensional $\{Q/M,a/M,\hbox{Im}(\omega M)\}$ plot, all these $\ell=m\neq2$ QNM are below the $Z_2$, $\ell=m=2$.

\begin{figure}[th]
\centering
\includegraphics[width=.38\textwidth]{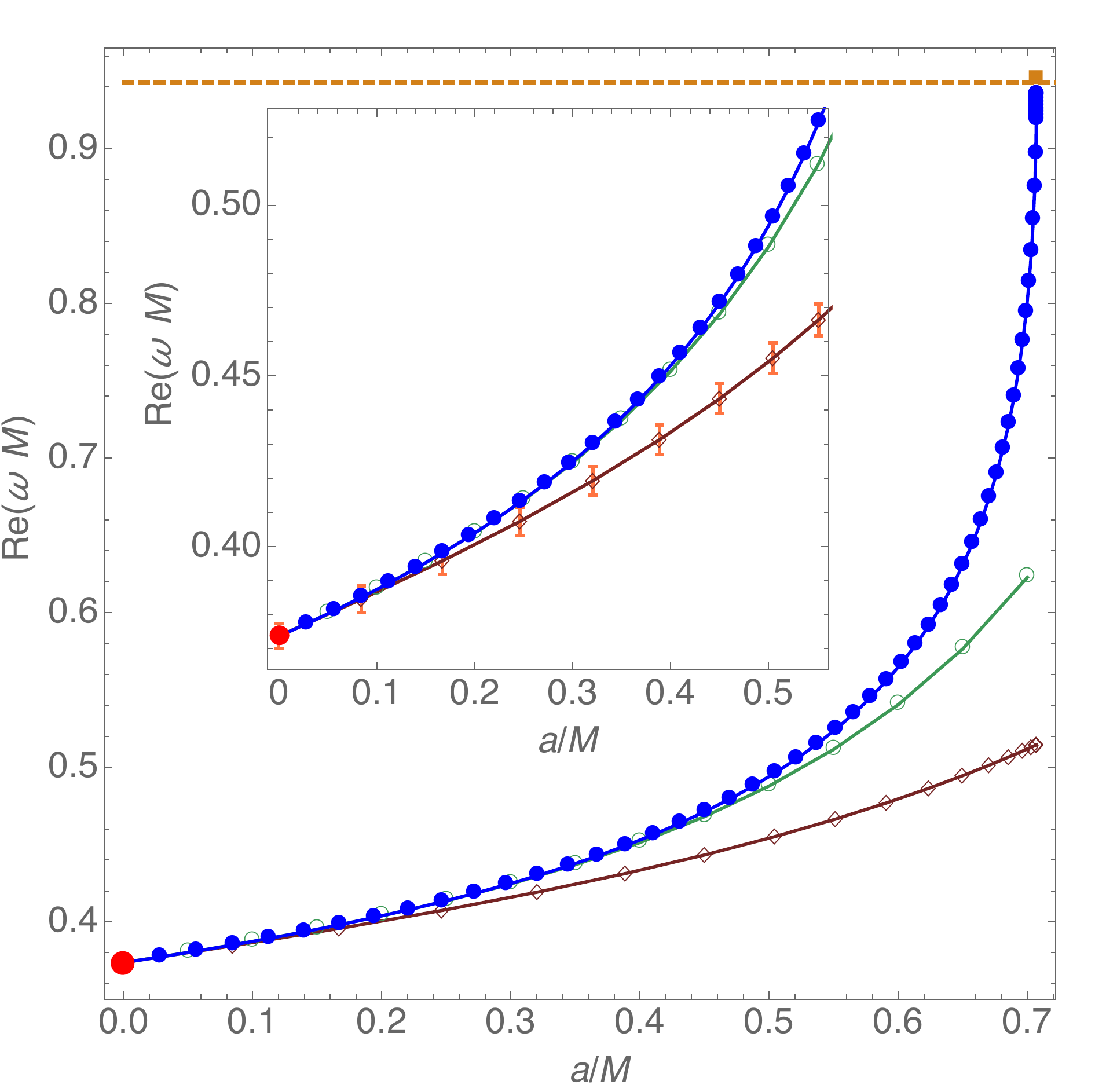}
\caption{Comparison (for Re$\,\omega$) between the exact $\ell=m=2$ $Z_2$ QNMs of KN with $Q=a$ (blue disks) with the small $a$ approximations of \cite{Pani:2013ija} (red diamonds with their $1\%$ error bar) and with the small $Q$ approximations of \cite{Mark:2014aja} (green circles).}\label{Fig:pertRe}
\end{figure}

\begin{figure}[th]
\centering
\includegraphics[width=.40\textwidth]{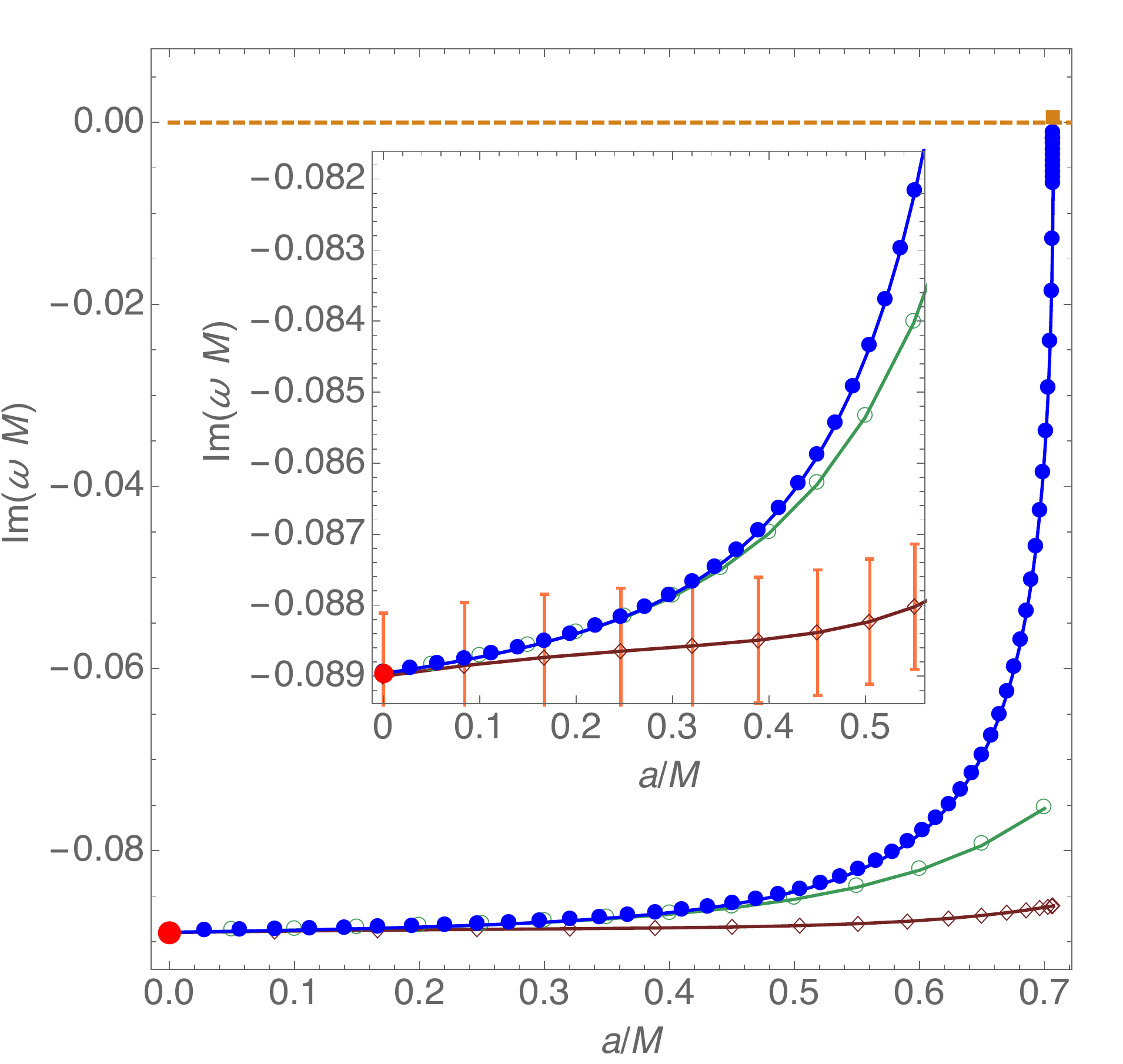}
\caption{Similar comparison as in Fig. \ref{Fig:pertRe} but now for Im$\,\omega$.}\label{Fig:pertIm}
\end{figure}

Previously, there were some attempts to find the QNMs of the KN BH using a perturbative analysis, notably for small $a$ around the RN QNMs \cite{Pani:2013ija,Pani:2013wsa} and for small $Q$ around the Kerr QNMs \cite{Mark:2014aja}. We use these perturbative results to further check our results for small $a$ or $Q$, thus establishing the regimes of validity of the aforementioned approximations. To compare the two perturbative analyses in a single graphic it is convenient to look at QNMs with $Q=a$, see Figs. \ref{Fig:pertRe} (real part) and \ref{Fig:pertIm} (imaginary part). We see that the approximations of \cite{Pani:2013ija,Pani:2013wsa} (\cite{Mark:2014aja}) are within $1\%$ of the exact results up to $\sim 25 \%$ ($\sim 70 \%$) of extremality.

\begin{figure}[th]
\centering
\includegraphics[width=.42\textwidth]{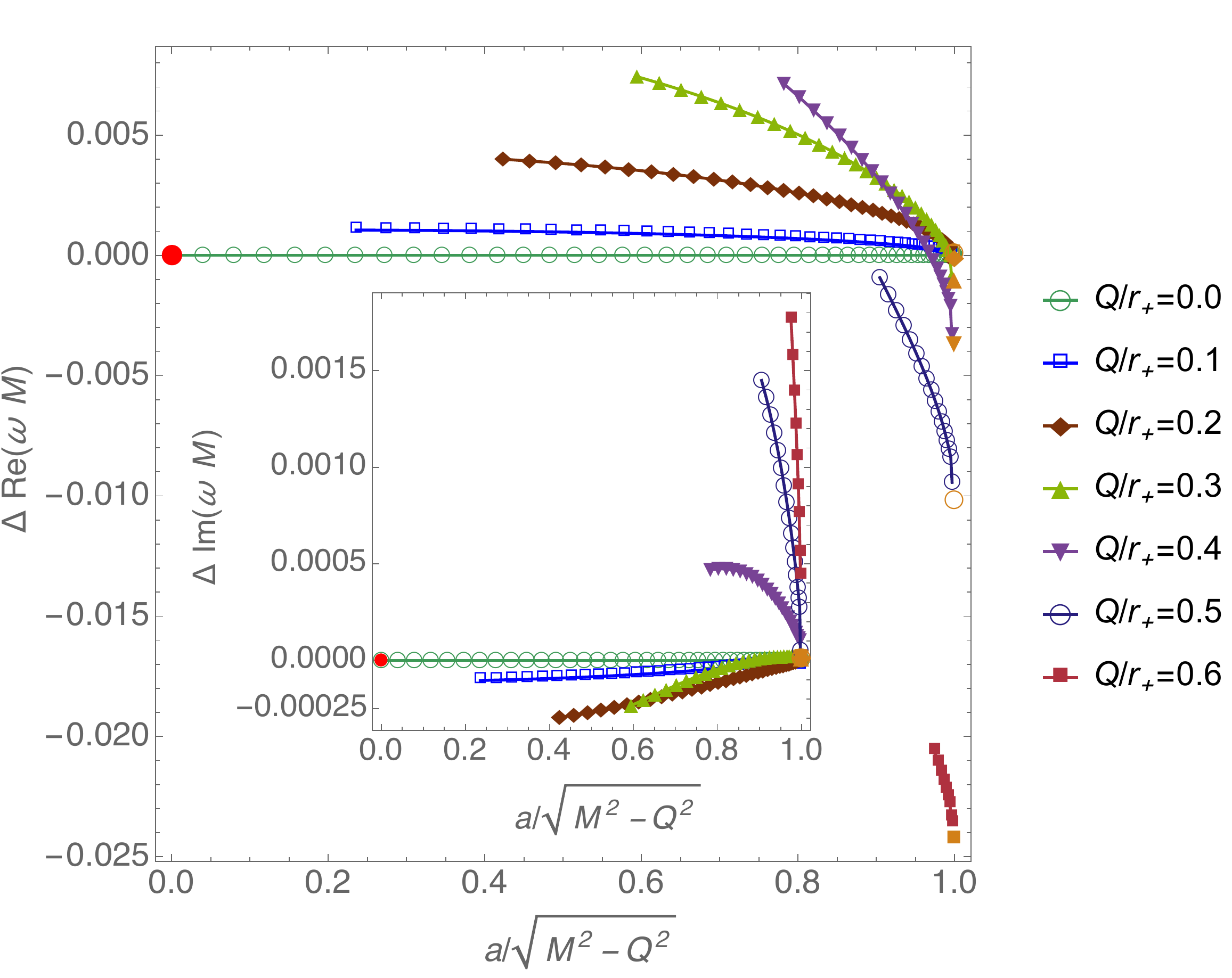}
\caption{Some of the $\ell=m=2$ QNMs of Figs. \ref{Fig:l2m2re} and \ref{Fig:l2m2im} but truncated for $a/Q>1$. They do not display the scaling conjectured in \cite{Zilhao:2014wqa}.}\label{Fig:scale}
\end{figure}

Ref. \cite{Zilhao:2014wqa} considered the full time evolution of some KN BHs with a given initial perturbation and found no sign of a nonlinear instability, which is consistent with our full parameter scan of the QNMs up to $a/a_{ext} = 0.99999$. Ref. \cite{Zilhao:2014wqa} also finds numerical evidence that some $\ell=m=2$ QNMs of a KN BH (with $a/Q>1$) should have the scaling $\omega=\omega\left(a/\sqrt{M^2-Q^2}\right)$. We can test this claim with {\it very high} accuracy and we find that it does not hold (although it must be emphasised that our linear results are well within the numerical accuracy of \cite{Zilhao:2014wqa}; they differ by at most 1\%).  Indeed, in Fig. \ref{Fig:scale} we pick some KN BHs with fixed $Q/r_+$ and $a/Q>1$. For $\ell=m=2$, we plot the difference $\Delta (\omega M)$ between $\omega M$ and the QNM frequency of the $Q=0$ KN BH as a function of $a/\sqrt{M^2-Q^2}$ (real part of $\omega M$ in the main plot and imaginary part in the inset plot). If the scaling proposed in \cite{Zilhao:2014wqa,Hod:2014uqa}  were present, all these curves should superpose at $\Delta (\omega M)=0$. This is not the case: e.g. $ |\Delta {\rm Im}(\omega M)|$ can be as high as $0.02$ (when the error in our results is $\pm 10^{-10}$).  Coming back to the fact that these modes approach $m\Omega_H^{ext} +0\, i$ at extremality, this had to be the case since $\Omega_H^{ext}\neq\Omega_H^{ext}\left(a/\sqrt{M^2-Q^2}\right)$ (see extremal points on Fig. \ref{Fig:scale}). To sum, although the proposed scaling fails to hold just by a small relative amount (less than 1\%), our numerical and analytical considerations show that it is only approximate, but not exact. 

Finally, we present a simple proof of the isospectral property \cite{Chandra:1983} of the Schw and RN QNMs \footnote{When $a=0$, (4) decouples and we can look independently to the gravitational $\psi_{-2}$ or  electromagnetic  $\psi_{-1}$ perturbations. In the main text the proof is for $\psi_{-2}$ in the Schw BH but it extends trivially to $\psi_{-1}$ modes and RN background.}. We can use the Teukolsky equation to study the QNMs of the Schw BH, instead of the Regge-Wheeler$-$Zerilli (RWZ) formalism. The two must give the same spectrum. The Teukolsky formulation has a single gauge invariant variable $\psi_{-2}$ that must translate into two gauge invariant variables in the RWZ formulation, namely Regge-Wheeler's  $\Phi_{\rm v}$ and Zerilli's $\Phi_{\rm s}$ eigenfunctions. Refs. \cite{Chandra:TekRWZ,Sasaki:1981,Dias:2013sdc} give the unique differential map that applied to $\psi_{-2}$ constructs $\Phi_{\rm v}$ and another that constructs $\Phi_{\rm s}$. Isospectrality is the statement that $\Phi_{\rm v}$ and $\Phi_{\rm s}$ have the same QNM spectrum. Since $\Phi_{\rm v}$ and $\Phi_{\rm s}$ are constructed from the same Teukolsky $\psi_{-2}$, it follows that this must be the case.  

%%%%%%%%%%%%%%%%%%%%%
\smallskip

%%%%%%%%%%%%%%%%%%%%%%%%%%%%%%%%%%%%%%%%%%%%%%%%%%%%%%%%%%%%%%%%%%%%%%%%%%
\noindent{\bf Acknowledgments.}
It is a pleasure to thank Vitor Cardoso, Luis Lehner and Miguel Zilh\~ao for discussions.
The authors thankfully acknowledge the computer resources, technical expertise, and assistance provided by CENTRA/IST. Some of the computations were performed at the cluster `Baltasar-Sete-S\'ois' and supported by the DyBHo-256667 ERC Starting Grant. OJCD acknowledges the kind hospitality of the Yukawa Institute for Theoretical Physics, where part of this work has been done during the workshop ``Holographic vistas on Gravity and Strings'', YITP-T-14-1. OJCD is supported by the STFC Ernest Rutherford grants ST/K005391/1 and ST/M004147/1. MG is supported by King's College, Cambridge. The research leading to these results has received funding from the European Research Council under the European Community's Seventh Framework Programme (FP7/2007-2013) / ERC grant agreement no. [247252].

%%%%%%%%%%%%%%%%%%%%%%%%%%%%%%%%%%%%%%%%%%%%%%%%%%%%
%%%%%%%%%%%%%%%%%%%%%%%%%%%%%%%%%%%%%%%%%%%%%%%%%%%%
%%% BEGINNING OF APPENDIX %%%
\begin{appendix}
%%%%%%%%%%%%%%
\section{Supplementary material: Derivation of the coupled equations \eqref{coupledeqns}}\label{sec:Appendix1}
%%%%%%%%%%%%%%%%%%%%%%%%%%%%%%%%%%%%%%%%%%%%%%%%%%%

The derivation of the two coupled PDEs \eqref{coupledeqns} is similar to the derivation of the Chandrasekhar coupled PDEs \cite{Chandra:1983} (\cite{Mark:2014aja}), except that we do not fix gauge.  We derive the equations using the NP formalism.  All the NP equations quoted in this section refer to equations in \cite{Stephani:2003tm}.  The first equation is derived by considering $\delta^*$(7.32d) $-$ $\Delta$(7.32c). The resulting expression is simplified by using these as well as other NP equations to express all perturbed quantities in terms of $\Psi_{4}^{(1)}$, $\Psi_{3}^{(1)}$ and $\Phi_{2}^{(1)}$.  In particular, NP eqn.\ (7.21j) is useful.  The derivation of the second equation is slightly less straightforward.  Here we consider an appropriate combination of $[ D\textup{(7.32d)} - \delta \textup{(7.32c)}]$ and $[ \Delta\textup{(7.23)} - \delta^* \textup{(7.25)}]$ that is suggested by the gauge invariant quantity that comes out of the first equation, i.e. we manipulate the NP equations in such a way as to obtain an expression in which the only second order in derivative terms are of the form $(D\Delta-\delta\delta^*)\varphi_{-1}$.  Subsequently, we simplify the resulting expression as before.  The fact that the Maxwell NP eqns.\ (7.23) and (7.25) mix background and perturbed quantities does not cause any problems as the background contributions cancel once the expression is fully simplified as described above. 

Finally, the differential operators $\{\mathcal{O},\mathcal{P},\mathcal{Q}\}$ introduced in \eqref{coupledeqns} are:
\begin{eqnarray} \label{def:opsOPQ}
\mathcal{O}_{-2} &=& (\Delta + 3\gamma - \gamma^* + 4\mu + \mu^*)(D-\rho) \notag \\
&& - (\delta^* + 3\alpha + \beta^* + 4\pi - \tau^*)(\delta + 4\beta -\tau) - 3\Psi_2, \notag \\
\mathcal{P}_{-2} &=& 2 - 4 (\Delta + 3\gamma -\gamma^*)A_{-} (D-\rho) \notag \\
&& -4 (\tau^* - \pi) A_{+} (\delta^* + 4 \beta - \tau), \notag \\
\mathcal{Q}_{-2} &=& 2/\Phi_{1}^{(0)}{}^{*} \big\lbrace (\Delta + 3\gamma - \gamma^* + 2\mu)A_{-}(\delta^* + 2\alpha + 6\pi) \notag \\
&& + (\tau^* - \pi)A_{+}(\Delta + 2\gamma + 6\mu) \big\rbrace , \notag \\
\mathcal{O}_{-1} &=& (\Delta + 3\gamma + \gamma^* + 5\mu + \mu^*)(D-4\rho)  \\
&& - (\delta^* + \alpha + \beta^* + 5\pi - \tau^*)(\delta + 2\beta -4\tau), \notag \\
\mathcal{P}_{-1} &=& 2 (D - 4\rho + \rho^*)A_{+} (\Delta + 2\gamma + 6\mu) \notag \\
&& + 2  (\delta + 3 \beta - \alpha^* - 4\tau - \pi^*)A_{-}(\delta^* + 2\alpha +6 \pi), \notag \\
\mathcal{Q}_{-1} &=& -4 \Phi_1^{(0)} \big\lbrace (D - 2\rho + \rho^*)A_{+}(\delta^* + 4\beta - \tau) \notag \\
&& + (\delta + 3\beta - \alpha^* - 2 \tau + \pi^*)A_{-}(\Delta + 4\gamma - \rho) \big\rbrace, \notag
\end{eqnarray}
where $A_{\pm}=(3\Psi_{2}^{(0)} \mp 2 \Phi_{11}^{(0)})^{-1}$.

%%%%%%%%%%%%%%
\section{Supplementary material: The second order differential operators $\{\mathcal{F},\mathcal{G},\mathcal{H}\}$}\label{sec:Appendix2}
%%%%%%%%%%%%%%%%%%%%%%%%%%%%%%%%%%%%%%%%%%%%%%%%%%%

The coupled system of two PDEs \eqref{ChandraEqs} that describe the most general linear perturbation of a KN BH (in the Newman-Penrose formalism) were first derived, in the phantom gauge $\Phi_{0}^{(1)}=\Phi_{1}^{(1)}=0$, by Chandrasekhar in his celebrated textbook \cite{Chandra:1983}. A nice compact summary of  this derivation can also be found in Appendix of \cite{Mark:2014aja}. This still yields our equations \eqref{ChandraEqs} but this time with the associated gauge fixing applied to the variables \eqref{gauging}.

The second order differential operators $\{\mathcal{F},\mathcal{G},\mathcal{H}\}$ that appear in \eqref{ChandraEqs} are explicitly given by
\begin{eqnarray}\label{def:opsFGH}
\mathcal{F}_{-2}&=&\Delta\DD_{-1}^\dagger\DD_0 +\LL_{-1}\LL_2^\dagger -6i \omega\bar{r} \,,
\nonumber \\
\mathcal{G}_{-2}&=&\Delta \DD_{-1}^\dagger \alpha_-\bar{r}^*\DD_0  -3\Delta \DD_{-1}^\dagger \alpha_- 
- \LL_{-1}\alpha_+ \bar{r}^* \LL_2 ^\dagger \nonumber \\
&& +3 \LL_{-1} \alpha_+ i a \sin \theta \,, 
\nonumber \\
\mathcal{H}_{-2}&=&-\Delta \DD_{-1}^\dagger \alpha_- \bar{r}^* \LL_{-1}  -3 \Delta \DD_{-1}^\dagger \alpha_- i a \sin \theta \nonumber \\
&&
-\LL_{-1} \alpha_+ \bar{r}^* \Delta \DD_{-1}^\dagger  -3\LL_{-1} \alpha_+ \Delta  \,,
\nonumber \\
\mathcal{F}_{-1}&=&\Delta\DD_1\DD_{-1}^\dagger +\LL_2^\dagger\LL_{-1}-6i \omega\bar{r} \,,
\\
\mathcal{G}_{-1}&=& - \DD_0 \alpha_+ \bar{r}^* \Delta \DD_{-1}^\dagger  -3 \DD_0 \alpha_+ \Delta 
 +\LL_2^\dagger \alpha_- \bar{r}^* \LL_{-1} \nonumber \\
&& +3 \LL_2^\dagger \alpha_- i a\sin\theta \,,
\nonumber \\
\mathcal{H}_{-1}&=& -\DD_0 \alpha_+ \bar{r}^* \LL_2^\dagger  +3 \DD_0 \alpha_+ i a \sin \theta  
 -\LL_2^\dagger \alpha_- \bar{r}^* \DD_0  \nonumber \\
&& +3 \LL_2^\dagger \alpha _- \,,
\nonumber
\end{eqnarray}
with $\bar{r}\equiv  r+ia\cos\theta$, $\alpha_\pm \equiv \left[3(\bar{r}^2M-\bar{r} Q^2)\pm Q^2\bar{r}^*\right]^{-1}$,  and we introduce the radial and angular Chandrasekhar operators,
\begin{eqnarray}\label{def:DL}
&& \DD_j = \partial_r+\frac{i K_r}{\Delta}+2j\frac{(r-M)}{\Delta}, \quad K_r=am-(r^2+a^2)\omega; \nonumber \\
 && \LL_j = \partial_\theta+K_{\theta}+j\cot\theta, \quad K_{\theta}=\frac{m}{\sin\theta}-a\omega\sin\theta. 
\end{eqnarray}
The complex conjugate of these operators, namely $\DD_j^\dagger $ and $\LL_j^\dagger$ can be obtained from $\DD_j$ and $\LL_j$ via the replacement $K_r \to - K_r$ and $K_{\theta} \to - K_{\theta}$, respectively.
% \quad \DD_j^\dagger \equiv \partial_r-\frac{i K_r}{\Delta}+2j\frac{(r-M)}{\Delta}
% \quad \LL_j^\dagger \equiv \partial_\theta-K_{\theta}+j\cot\theta

%%%%%%%%%%%
\end{appendix}
%%%%%%%%%%%%%%%%%%%%%%%%%%%%%%%%%%%%%%%%%%%%%%%%%%%%

%%%%%%%%%%%%%%%%%%%%%%%%%%%%%%%%%%%%%%%%%%%%%%%
\bibliography{refs}{}
\bibliographystyle{JHEP}

\end{document}